# Note: Eliminating Stripe Artifacts in Light-Sheet Fluorescence Imaging

S. M. Salili, M. Harrington, D. J. Durian

*Department of Physics & Astronomy, University of Pennsylvania, Philadelphia, Pennsylvania 19104, USA*

We report two techniques to mitigate stripe artifacts in light-sheet fluorescence imaging. The first uses an image processing algorithm called the multidirectional stripe remover (MDSR) method to filter stripes from an existing image. The second uses an elliptical holographic diffuser (EHD) with strong scattering anisotropy to prevent stripe formation during image acquisition. These techniques facilitate accurate interpretation of image data, especially in denser samples. They are also facile and cost-effective.

Light-sheet fluorescence imaging is a robust technique that is gaining popularity and wider use in a variety of systems. For living biological samples ranging from embryonic development studies all the way to neuroscience, it provides high resolution, contrast, specificity, penetration depth, and image acquisition speed, and has relatively low levels of phototoxicity and photobleaching[1–5]. For colloidal and granular materials, it allows particles to be tracked for study of soil mechanics[6,7], flow behavior and rheology[8–23], packing and jamming phenomena[7,21,22,24,25]. For all such applications, a sample is illuminated by a laser sheet, resulting in a two dimensional cross-sectional image, which can be combined into a stack of scanned images to obtain a three-dimensional representation[26,27]. The optical access to the bulk of a colloidal or granular medium is accomplished by matching the refractive indices in a fluorescent dispersion of fluid and particles[26,28,29].

Although light-sheet fluorescence imaging provides essential structural information, the images are not pristine and often contains stripes (e.g. see Figure 1(a)). Such artifacts generically arise from either absorbing or scattering structures along the illumination light path[30–32]. A general ray tracing simulation is performed to highlight mismatch in the index of refraction of an object against a fluid, as the salient cause for formation of stripe artifacts. Additionally, it contributes a base upon which the experimental conditions can be fine-tuned to alleviate formation of stripe artifacts (see details and ray tracing code in supplementary material). For particulate suspensions, stripes can be lessened by improved refractive index matching; however, even a refractive index difference ($\Delta n = n_p - n_f$) as small as 0.001 can cause noticeable artifacts (see Figure 1(c-f)). This makes image analysis laborious and often inaccurate, leading to misinterpretation of the data. For example, stripes can cause standard particle-tracking algorithms[33,34] to erroneously report a series of adjacent particles.

Different approaches have been proposed heretofore by biologists to overcome the artifact issue. Several image processing algorithms for denoising the artifact-rich images have been developed[35–37]. Multidirectional light-sheet illumination (illuminating from different angle)[38,39], an airy beam[40], a Bessel beam[41,42] and a scanner beam[3], have all lessened the artifacts optically at their origin. By contrast, in granular matter physics, not much effort has been made to resolve this issue. Only recently, Houssais *et al.* and Dijksman *et al.* have proposed image processing algorithms to further destripe the images[7,31].

Unfortunately, existing stripe elimination methods often require sophisticated instrumentation and at the same time are not able to completely remove stripes in all directions. Therefore, here we describe two further techniques. The first one is a recent but not yet widely known image processing algorithm called the multidirectional stripe remover (MDSR) method created by Liang *et al.*[30]. Although MDSR has already been implemented for biological samples[30], here we implement it on granular materials and we provide the full MDSR code. The second technique is a novel passive optical device for creating multidirectional illumination using an elliptical holographic diffuser (EHD).

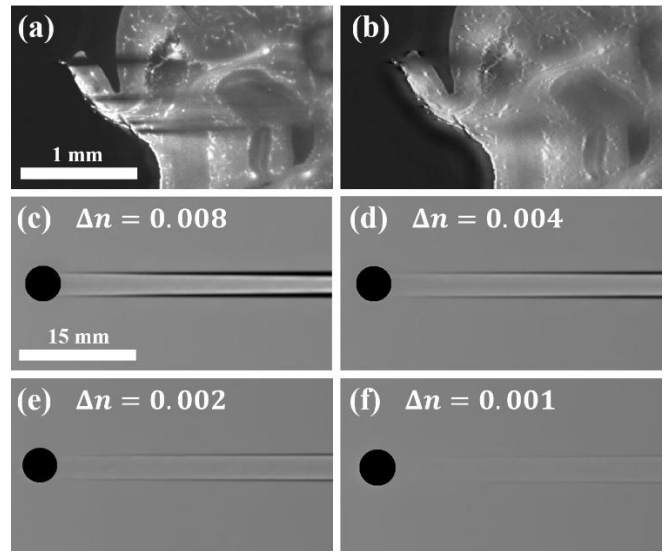

FIG. 1. Light-sheet fluorescence images of mouse embryo, (a) raw[43] with dark stripes due to upstream air bubbles and (b) destriped via multidirectional stripe remover (MDSR) method. (c-f) Ray tracing simulation results for a 4.8 mm diameter polymethylmethacrylate bead (PMMA, $n_p = 1.490 + \Delta n$) suspended in Triton X-100 solution ($n_f = 1.490$) with various refractive index differences.

To begin with the first method, the multidirectional stripe remover (MDSR) mitigates multidirectional stripe artifacts in an image using the combination of nonsubsampled contourlet transform[44,45] and fast Fourier transform filtering. MDSR results are controlled by user choice for the following five input parameters: $n_i$ (number of layers), $n_d$, (power of directional decomposition), $\sigma$ (controlling suppression degree), $\sigma_a$ (controlling suppression weight) and $\theta_{elim}$ (the angle or angles at which the stripes should be suppressed). Details are given by Liang et al.[30]. Directions for how to choose these parameters are included in our MATLAB implementation (code available in supplementary material).

As an example of the MDSR method, we apply it to the mouse embryo image of Figure 1(a), taken using a Zeiss light-sheet microscope[43]. Input parameters of $n_i = 5$, $n_d = 3$, $\sigma = 10$, $\sigma_a = 8$ and $\theta_{elim} = 0°$ were optimized by trial-and-error, giving the destriped image in Figure 1(b). While the severity of the black stripes is considerably lessened, there remains a noticeable diffuse darkening that still mars the image.

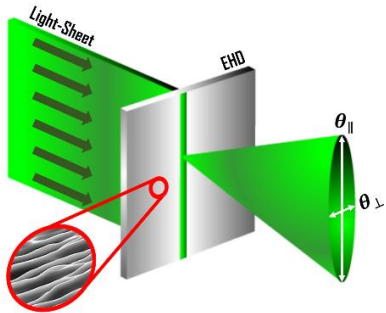

FIG. 2. Schematic of an elliptical holographic diffuser (EHD) that passively generates multidirectional illumination with diffusing anisotropy of $\theta_\parallel$ and $\theta_\perp$. The zoomed-in area shows a scanning electron micrograph of the EHD[46].

For the second method, an elliptical holographic diffuser is placed into the incident light-sheet, just in front of the sample, to prevent stripe formation in the first place. The principle is multidirectional illumination, as depicted in Figure 2. When illuminated with a narrow beam of light, an ordinary (circular) diffuser causes transmitted light to emerge isotropically at all angles. By contrast, an EHD causes transmitted light to spread very anisotropically: by up to a large amount $\theta_\parallel = O(45°)$ in one direction, and by only a small amount $\theta_\perp = O(0.1°)$ orthogonally. Therefore, a thin light-sheet of parallel rays is transformed into a similarly-thin diffuse light-sheet with rays going at a wide range of angles, $-\theta_\parallel/2$ to $+\theta_\parallel/2$. The resulting multidirectional illumination prevents an absorbing particle from casting a sharp shadow, and thus prevents stripes in general.

Elliptical Holographic diffusers consist of a thin sheet, one side of which is replicated from a holographic recorded master, producing a texturized surface structure[47]. The surface pattern is pseudo-random, non-periodic and resembles a micron-sized sand dune with hillocks and troughs[48] that have long-range orientational order (see Figure 2) and cause the light to spread in a plane without Bragg peaks[49]. They are commercially available, e.g. from Edmund Optics or Luminit.

Primary uses are in the liquid crystal display industry, to eliminate the Moiré pattern, increase backlight brightness, and modify its viewing cone[50]; in machine vision, to provide the necessary uniformity in line scan metrology[51]; and generally in any system using simple ellipsoidal optics, to smooth out the hot spots and homogenize lighting[52].

For maximum artifact removal, larger $\theta_\parallel$ and smaller $\theta_\perp$ are generally better. The degree of artifact removal for a larger $\theta_\parallel$, can be verified by changing $\theta_\parallel$, using our ray tracing simulations. For a typical system like ours, $\theta_\parallel = 30°$, 40° or 60° all perform similarly well. The value of $\theta_\perp$ need not be smaller than the divergence of the laser sheet thickness. But it should be small enough that the sheet does not significantly thicken as it traverses the sample. For our system, $\theta_\perp = 0.2°$ performs well, since the diffuse light sheet does not become thicker than our particles size; $\theta_\perp = 1°$ performs nearly as well for our 4.8-mm particles (details on thickness of diffuse light-sheet available in supplementary material).

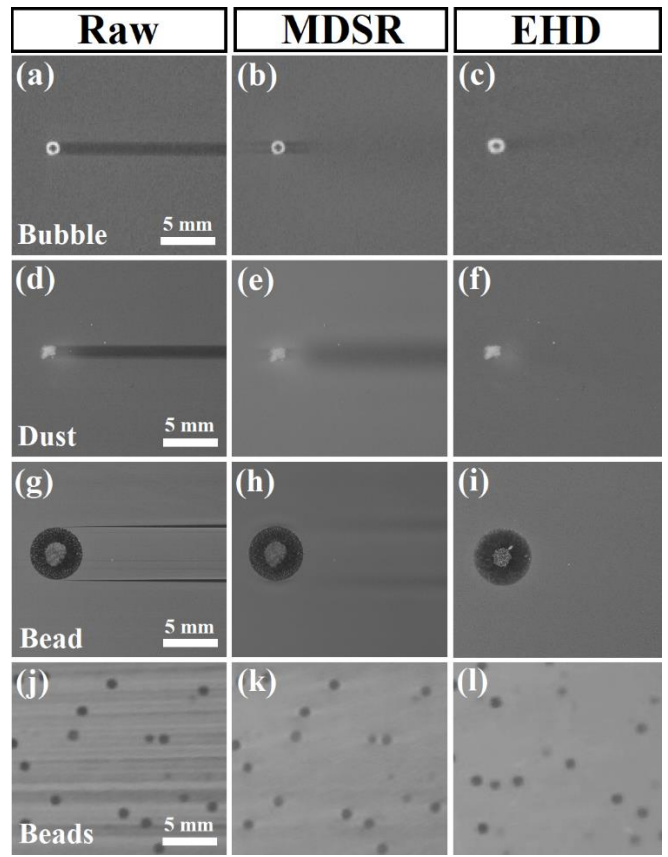

FIG. 3. Raw images of various objects suspended in Triton X-100 index-matching solution (left column) destriped via multidirectional stripe remover method (MDSR, middle column). Raw images of same systems illuminated through an elliptical holographic diffuser with diffusing angles of $\theta_\parallel =40°$ and $\theta_\perp =0.2°$ (EHD, right column).

For controlled experimental tests and comparison of the two stripe-removal methods, we use a refractive index (RI)-matched system of Triton X-100 (Sigma-Aldrich, nominal RI of $n_f = 1.49$) and PMMA particles (Engineering Laboratories, nominal RI of $n_f = 1.49$, diameters $d$=1.0 or

4.8 mm) in a rectangular acrylic vessel ($30 \times 4 \times 4$ cm$^3$). We estimate the refractive index mismatch to be $\Delta n = 0.002$. Fluorescent dye (1 μM, Exciton, pyrromethene 597) is dissolved in the suspension and illuminated with a green laser sheet (Coherent StingRay, $\lambda$ =517 nm, 50 mW, fan angle = 45°, thickness ≃ 0.1 or 0.5 mm). Consequently, the particles appear as dark circles, and are imaged with a Nikon D90 DSLR camera equipped with a 550-nm high performance longpass filter (Edmund Optics). The elliptical holographic diffuser used for the images discussed below is made from polycarbonate (Edmund Optics, $\theta_{\parallel}$ =40°, $\theta_{\perp}$ =0.2°); it transmits more than 85% of incident light.

Figure 3 shows images of an air bubble, a dust particle, a large indexed-matched PMMA bead, and a suspension of small index-matched beads, all in the same fluid. Raw images with traditional light-sheet illumination, in the left column, all exhibit obvious stripe artifacts because the particles block the downstream fluid from being uniformly illuminated. The results of filtering out these stripes by post-processing with MDSR are shown in the middle column. This vastly improves image quality; however, just as seen in Figure 1(b), artifacts are not fully eliminated. Raw images from illuminating the samples through an EHD placed on the face of the sample cell are shown in the right column. As seen, the resulting multi-angle light-sheet illumination successfully prevents the formation of stripes. Only slight artifacts occur close behind the gas bubble and the dust particle; no artifacts are evident for the PMMA beads. Comparison of image quality for the middle and right columns highlights the advantages of EHD over MDSR (see quantitative comparison in supplementary material). Both methods successfully mitigate stripe artifacts; however, EHD delivers superior performance accompanied by simpler implementation.

We thank Andrea Stout, Michal Shoshkes-Carmel and Klaus Kaestner for providing the sample image of Figure 1(a). This work was supported by the NSF through grant DMR-1305199.

## SUPPLEMENTARY MATERIAL

See supplementary material for our MDSR and ray tracing simulations codes, both written in MATLAB.